\documentclass{osa-article}

\journal{osajournal}

\articletype{Research Article}

\begin{document}

\title{Mid-infrared frequency combs at 10 GHz}

\author{Abijith S. Kowligy\authormark{1,2,*}, David R. Carlson\authormark{1}, Daniel~D.~Hickstein\authormark{1}, Henry~Timmers\authormark{1}, Alexander~J.~Lind\authormark{1,2}, Peter G. Schunemann\authormark{1,2}, Scott~B.~Papp\authormark{1,2}, Scott~A.~Diddams\authormark{1,2}}

\address{\authormark{1}Time and Frequency Division, National Institute of Standards and Technology, Boulder, CO 80305 USA}
\address{\authormark{2}Department of Physics, University of Colorado, Boulder, CO 80309 USA}
\address{\authormark{3}BAE Systems, Nashua, NH USA}
    
\email{\authormark{*}abijith.kowligy@gmail.com}

\begin{abstract}
We demonstrate mid-infrared (MIR) frequency combs at 10 GHz repetition rate via intra-pulse difference-frequency generation (DFG) in quasi-phase-matched nonlinear media. Few-cycle pump pulses ($\mathbf{\lesssim}$15~fs, 100~pJ)  from a near-infrared (NIR) electro-optic frequency comb are provided via nonlinear soliton-like compression in photonic-chip silicon-nitride waveguides. Subsequent intra-pulse DFG in periodically-poled lithium niobate waveguides yields  MIR frequency combs in the 3.1--4.8~\textmu{}m region, while orientation-patterned gallium phosphide provides coverage across 7--11~\textmu{}m. Cascaded second-order nonlinearities simultaneously provide access to the  carrier-envelope-offset frequency of the pump source via in-line \textit{f-2f} nonlinear interferometry. The high-repetition rate MIR frequency combs introduced here can be used for condensed phase spectroscopy and applications such as laser heterodyne radiometry.
\end{abstract}

Coherent broadband frequency comb sources in the mid-infrared (MIR, 3--25 \textmu{}m) are valuable for studying light--matter interactions in molecular systems, e.g. trace-gas sensing and precision spectroscopy\cite{cossel_gas-phase_2017,ycas_high-coherence_2018,muraviev_massively_2018,Changala2019}, time-resolved kinetics and dynamics of chemical reactions\cite{fleisher_mid-infrared_2014,klocke_single-shot_2018}, and mapping the secondary structure of bio-molecular compounds \cite{Kowligy2019, Weichmann2019}. Because only a few broad bandwidth lasers directly emit radiation in the mid-infrared \cite{Vasilyev:19,Duval:15,faist_quantum_2016,Antipov2016}, nonlinear frequency conversion is often employed to generate frequency combs in this spectral region \cite{schliesser_mid-infrared_2012,cossel_gas-phase_2017}. Examples include difference frequency generation (DFG) \cite{erny_mid-infrared_2007,Gambetta:08,sell_field-resolved_2008,cruz_mid-infrared_2015,ycas_high-coherence_2018},  $\chi^{(2)}$\cite{Adler2009,Leindecker:11,Marandi:16,vainio_mid-infrared_2016,maidment_long-wave_2018,muraviev_massively_2018} and $\chi^{(3)}$\cite{mirmicrocombs16,wang_mid-infrared_2013} parametric oscillators, as well as supercontinuum sources  \cite{petersen_mid-infrared_2014,kuyken_-chip_2011,Hudson:17,Nader:19}. With intrapulse DFG, as we employ here, the broad spectrum of a near-infrared pulse is downconverted to the MIR~\cite{huber2000generation,keilmann_time-domain_2004,pupeza_high-power_2015,butler2019watt,timmers_molecular_2018,Vasilyev:19}. This is an attractive alternative to $\chi^{(2)}$ optical parametric oscillators or the more typical two-branch DFG, because it avoids the need for spatial and temporal overlap of multiple beams and it provides broad simultaneous MIR spectral coverage.  Recently, intrapulse DFG spectra from 3--12 \textmu{}m have been demonstrated with combined brightness and bandwidth that is otherwise only available with a synchrotron light source \cite{timmers_molecular_2018,pupeza_high-power_2015,Kowligy2019,seidel_multi-watt_2018}.

Across these nonlinear techniques, most MIR frequency combs operate at rates on the order of 100 MHz due to the required nanojoule (or greater) pulse energies. For MIR spectroscopy of small gas-phase molecules at ambient temperatures and pressures, the $\sim$100 MHz comb mode spacing is a fraction of the typical linewidths.  However, for larger gas-phase molecules, solid-state, or liquid samples, the typical 100 MHz mode spacing is too fine, and sampling of a spectral resonance at 10 GHz (0.3 cm$^{-1}$) or greater would be more appropriate. Such large mode-spacing ($\geq$10~GHz) in the MIR is also highly desirable for applications such as imaging of biological materials \cite{tu2016stain,hermes2018mid,seddon2018prospective} and spectroscopy of novel quantum materials \cite{muller_quantum_2018} that demand fast acquisition rates ($\geq$1~MHz) and broad spectral coverage.

 In this work, we generate broad bandwidth and tunable MIR frequency combs at a rate of 10 GHz using common-place telecom-wavelength components and integrated photonics. We leverage a robust electro-optic (EO) comb generator \cite{carlson_ultrafast_2017} and $\chi^{(3)}$ spectral broadening in dispersion-engineered silicon nitride (Si$_3$N$_4$) waveguides to produce few-cycle near-IR pulses. This source then drives intrapulse DFG \cite{timmers_molecular_2018} in quadratic nonlinear materials such as periodically poled lithium niobate (PPLN) waveguides and orientation-patterned gallium phosphide (OP-GaP). With this combination of techniques, devices and materials, we overcome the challenging pulse energy limitations associated with high repetition rate sources to generate frequency combs across 3.1--4.8~\textmu{}m and 7--11~\textmu{}m.   In this regime, a few MIR frequency comb systems have been explored with mode-spacings ranging from a few gigahertz \cite{mayer_offset-free_2016,Rockmore:19} to 10 GHz and greater \cite{yu2018silicon,scalari2019chip}. In comparison, our approach provides frequency combs with broader and more widely tunable bandwidth than existing quantum cascade laser (QCL) combs \cite{faist_quantum_2016} and prior MIR electro-optic frequency comb demonstrations \cite{Yan2017,Jerez2018}. At the same time the 10 GHz mode spacing enables applications such as dual-comb spectroscopy \cite{coddington_dual-comb_2016} over a 10 THz spectral window with a 10 MHz acquisition rate, which will be ideal for the hyper-spectral imaging of bio-chemical systems and real-time tracking of their dynamics.

\begin{figure}[!t]
    \centering
    \includegraphics[width=\linewidth]{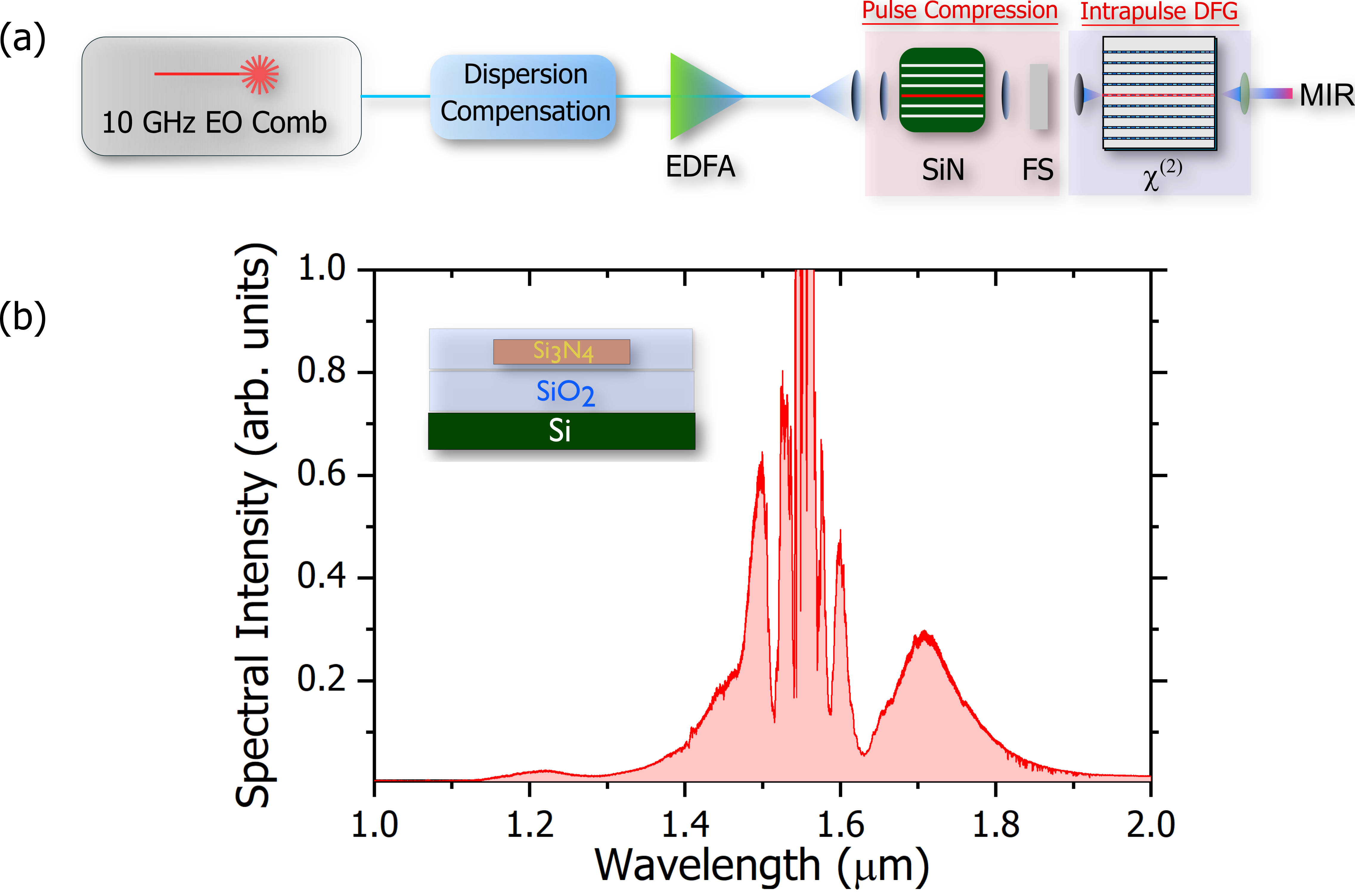}
    \caption{Mid-IR electro-optic comb experimental setup. (a) The experimental setup for generating 10 GHz mid-infrared frequency combs is shown. An electro-optic comb with 10 GHz repetition rate, centered at 1550 nm, is dispersion managed using a waveshaper and sent to an erbium-doped fiber amplifier (EDFA). Subsequent nonlinear spectral broadening in silicon nitride (SiN/Si$_3$N$_4$) nanophotonic waveguides results in few-cycle pulses that are approximately $15$ fs in duration with 100 pJ of pulse energy. The few-cycle pulse drives intrapulse DFG in $\chi^{(2)}$ nonlinear media to generate mid-IR combs. (b) The corresponding spectrum of the few-cycle pulse is shown. (Inset: the nanophotonic SiN waveguide geometry is shown). }
    \label{fig:setup}
\end{figure}

 We utilize a recently demonstrated EO frequency comb to generate few-cycle ($\lesssim$15 fs) pulses at 10 GHz in the near-infrared, centered at 1.55 \textmu{}m \cite{carlson_ultrafast_2017}. The EO comb is generated via cascaded phase- and amplitude-modulation (\cite{carlson_ultrafast_2017},Fig.~\ref{fig:setup}(a)). The output is amplified in an erbium-doped fiber amplifier to 4~W, spectrally broadened in 5-m long normal group-velocity dispersion highly nonlinear fiber (ND-HNLF) with dispersion parameters, $\beta_2$=~1.66 ps$^2$/km, $\beta_3$ = 9.76$x10^{-3}$ ps$^3$/km, and nonlinear coefficient $\gamma = 11$/W-km. The output of the ND-HNLF is compressed with free-space gratings to yield $\sim$100~fs pulses. The ND HNLF provides controlled spectral broadening that mitigates modulation instability, a common source for degradation of coherence in highly nonlinear spectral broadening processes. These pulses drive supercontinuum generation (Fig.~\ref{fig:setup}(b)) in 2-cm-long Si$_3$N$_4$ waveguides (800~nm thickness, widths ranging from 1200 nm to 2000 nm, coupling efficiency $\approx 75$\%) that exhibit anomalous dispersion at 1.55 \textmu{}m. Utilizing soliton self-compression in the waveguide \cite{carlson2019generating}, we generate few-cycle pulses with 1~W of average power, i.e. 100-pJ pulse energy, which are used to drive intrapulse DFG in PPLN or OP-GaP in the subsequent step.  
 
\begin{figure}[!t]
    \centering
    \includegraphics[width=\linewidth]{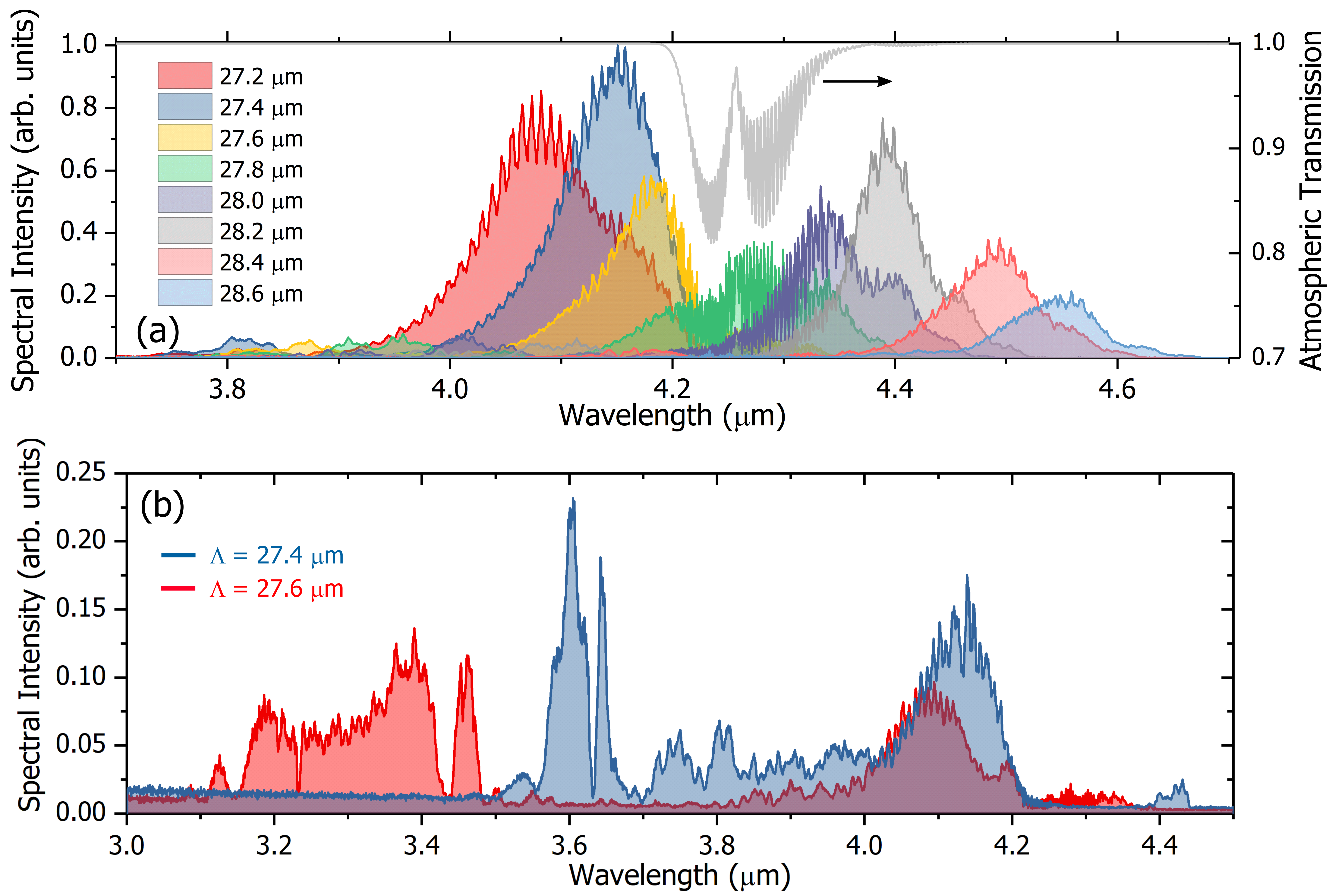}
    \caption{Frequency comb generation from 3.1--4.8 \textmu{}m. (a) Tunable MIR spectra are generated in 4-cm-long PPLN waveguides with various grating periods. Attenuation due to atmospheric carbon dioxide around 4.3 \textmu{}m is observed over 2 m propagation. (b) By pre-chirping the few-cycle pump pulse using bulk fused silica windows (1 cm thick), broadband frequency conversion is observed in the waveguides.}
    \label{fig:ppln}
\end{figure}

In one configuration, we employ 4-cm-long PPLN ridge waveguides with poling periods spanning 25.6--30 \textmu{}m to generate tunable, broadband combs from 3.8--4.6~\textmu{}m via intra-pulse DFG using only 50 pJ of in-coupled pump-pulse energy (Fig.~\ref{fig:ppln}(a)). The waveguides were fabricated by NTT Electronics America, by dicing Zn-doped PPLN waveguides on a lithium tantalate substrate \cite{asobe_high-power_2008}. The ridge dimensions are 12 $x$ 12 $\mu$m$^{2}$. The waveguide facets are anti-reflection coated for the pump wavelengths in the near-infrared.  The pump light is in-coupled using an 11-mm focal length aspheric lens and the waveguide output is collimated using a silver off-axis parabolic mirror. The input-output coupling efficiency is approximately 70\% for the pump light in this configuration. 

The center frequencies of the MIR output around 4.1~\textmu{}m ($\approx$73~THz) is understood by studying the spectral distribution of the pump (Fig.~\ref{fig:setup}(b)). The frequency difference between the two dominant peaks at the extrema, located at 1.2~\textmu{}m ($\approx$250~THz) and 1.7~\textmu{}m ($\approx$176~THz), yields 74~THz that corresponds to 4.05 \textmu{}m. We note that the 3-mm-thick Ge-substrate interference filters that transmit the MIR spectra form etalons and result in spectral modulation (Fig.~\ref{fig:ppln}(a)). The powers in the MIR range from 50~\textmu{}W for the output from the waveguide with $\Lambda$ =~ 28.6~\textmu{}m to 100~\textmu{}W for the output from the waveguide with $\Lambda$ =~27.2 \textmu{}m. Utilizing a 1-cm-long fused silica (FS) window to optimize the pump-pulse compression, we also observed broad spectra spanning 3.1--4.3~\textmu{}m that changed with the poling period (Fig.~\ref{fig:ppln}(b)). The mid-IR output was spectrally isolated using germanium filters and directed to a Fourier-transform spectrometer (FTS) and a liquid nitrogen-cooled mercury-cadmium-telluride (MCT) photodetector for analysis. The integrated output power was between 50--100~\textmu{}W, corresponding to $>10$ nW/nm spectral intensities.

\begin{figure}[!b]
    \centering
    \includegraphics[width=\linewidth]{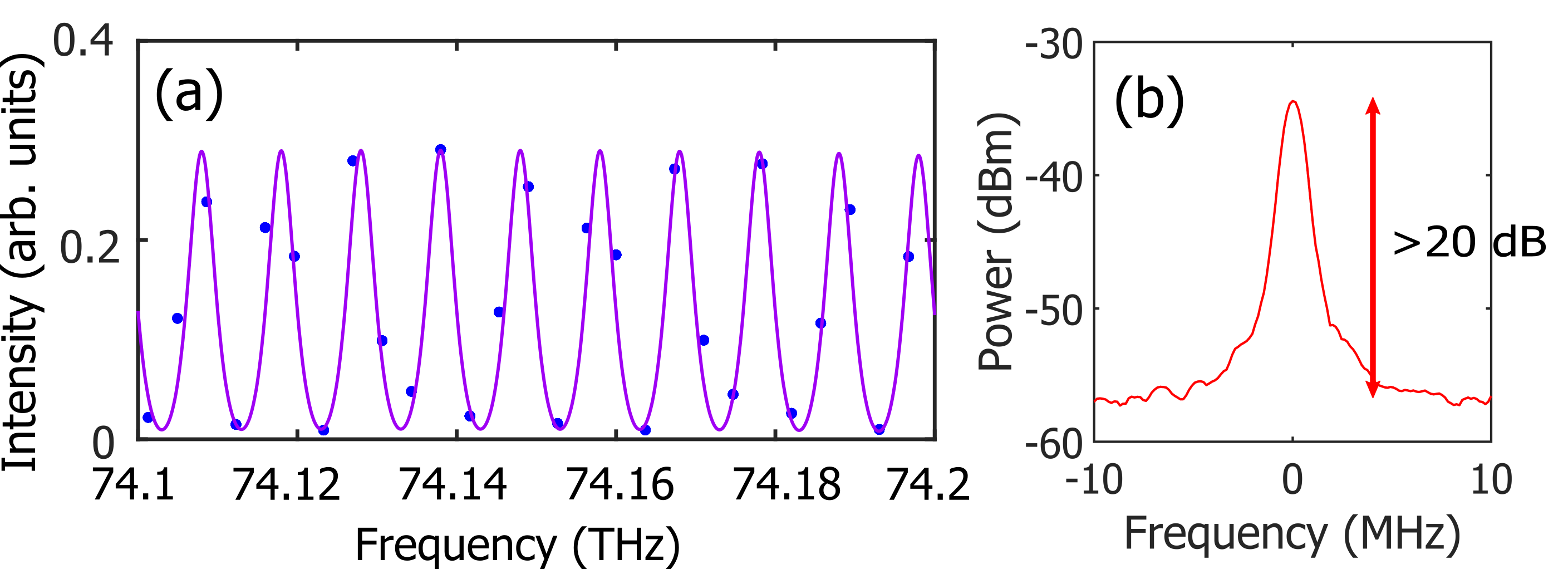}
    \caption{10 GHz modes in the mid-IR. (a) 10 GHz mode structure is observed in the mid-infrared around 74~THz ($\approx$4.04~\textmu{}m) as indicated by the blue dots and a fit (purple solid line). (b)~The carrier-envelope offset frequency for the NIR EO-comb is detected around 1.1 \textmu{}m at the output of the 4-cm-long PPLN waveguide on a high-speed silicon photodetector, owing to in-line $f$--$2f$ interferometry. The detection frequency is around 3.5 GHz and the resolution bandwidth is 1 MHz. }
    \label{fig:combmodes}
\end{figure}

The FTS can resolve individual comb-lines in the mid-IR (Fig.~\ref{fig:combmodes}(a)), showing that the nonlinear processes preserve the mode structure of the EO comb. However, the instrument resolution of 4 GHz artificially broadens the expected MHz-level comb tooth linewidths to the GHz-level. As a result, we use a numerical model, which is a sum of Lorentzian lineshapes at a repetition rate of 10 GHz, to fit the data in Fig.~\ref{fig:combmodes}(a). The model accounts for the finite resolution of the FTS and shows good agreement with the data. Moreover, the coherence of the comb in the NIR is manifest in the realization of an inline $f$-$2f$ interferometer in the PPLN waveguides. 

Owing to the strong nonlinearity and phase-matched 2~\textmu{}m to 1~\textmu{}m SHG \cite{iwakuni_generation_2016}, the carrier-envelope offset frequency, $f_{CEO}$, of the comb is detected at 1~\textmu{}m with $>$ 20 dB signal-to-noise ratio (SNR), as shown in the inset of Fig.~\ref{fig:combmodes}(b). We note here that the $f_{CEO}$ beat is strongly dependent on the output from the SiN waveguide as well as group-velocity matching in the 4-cm-long waveguide. In our experiment, the beat-note can be observed in all the waveguides generating MIR light. Prior experiments have stabilized this degree-of-freedom in the EO comb \cite{carlson_ultrafast_2017}, and the present experiment shows a simple route to its detection.

\begin{figure}[!t]
    \centering
    \includegraphics[width=\linewidth]{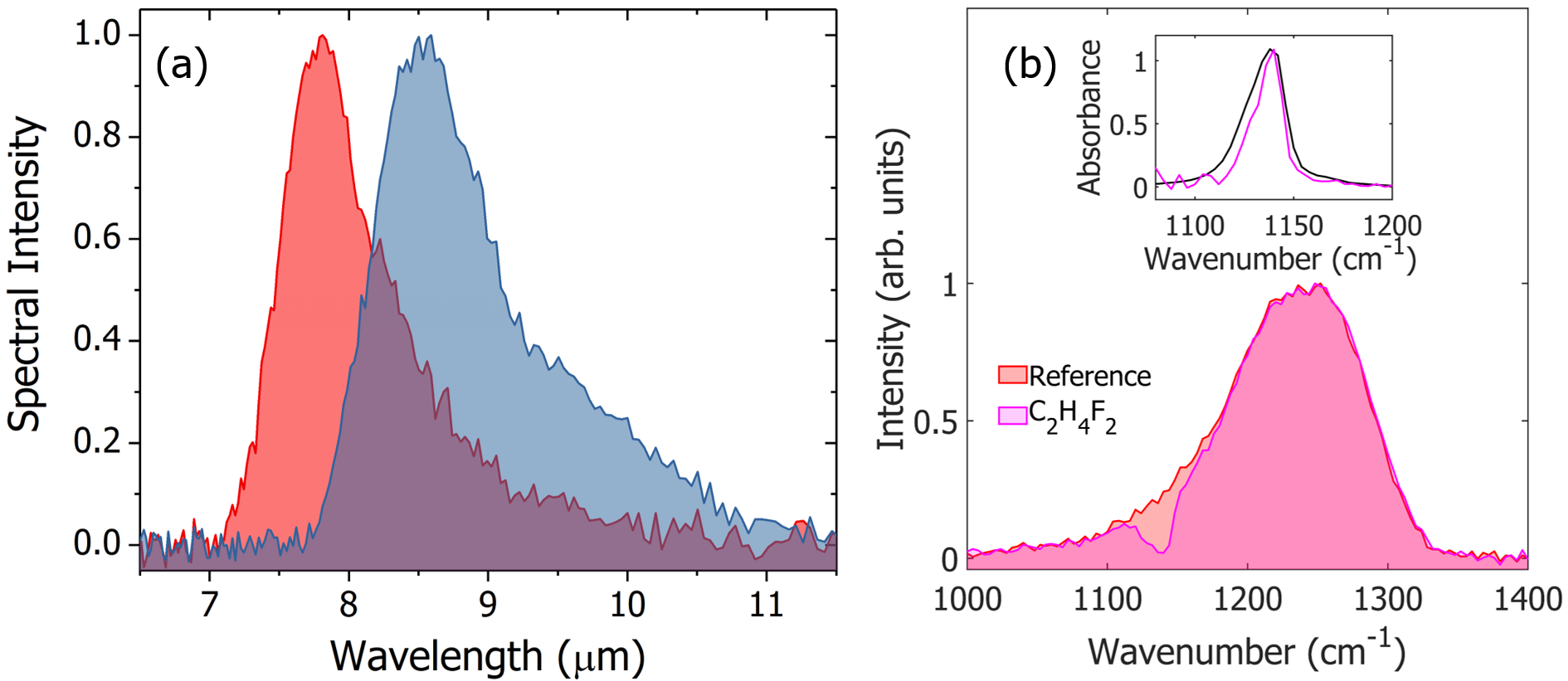}
    \caption{10 GHz frequency combs in the long-wave infrared. (a)~LWIR spectra from OP-GaP pumped by few-cycle pulses from 1280~nm (red) and 1620~nm (blue) width Si$_3$N$_4$ waveguides.  (b) The reference and transmission spectra for the measurement of C$_2$H$_4$F$_2$.  Inset: measured absorption spectrum compared with the NIST WebBook database \cite{nist_webbook}. }
    \label{fig:lwir}
\end{figure}

In addition to coherent sources in the 3--5~\textmu{}m window, frequency combs in the molecular fingerprint region (6--20~\textmu{}m) are of significant interest to the chemical, biological, and physical sciences \cite{schliesser_mid-infrared_2012,timmers_molecular_2018}. Using the same intra-pulse DFG technique, we extend the spectral coverage from 7--11~\textmu{}m using a 2-mm long bulk, uncoated OP-GaP crystal with poling period $\Lambda = 60$~\textmu{}m (Fig.~\ref{fig:lwir}(a)). OP-GaP provides a large second-order nonlinear coefficient ($\approx$35 pm/V) as well as a broad transparency  (0.6--12~\textmu{}m) into the long-wave infrared that is required for frequency comb generation in the molecular fingerprint region. 

With 800 mW of input pump power, the measured output powers range from 70--100~\textmu{}W. The output power of the OP-GaP was calibrated using the responsivity of the MCT photodetector at 7.5 \textmu{}m and 8.5 \textmu{}m, based on the average photocurrent measured. Accounting for the relatively high Fresnel losses at the crystal interface, we estimate that 300~\textmu{}W was generated. The integrated output power was calibrated using the MCT and a 4 cm$^{-1}$-resolution FTS recorded the spectrum. For the FTS measurements, we collect a background spectrum with the MIR comb blocked and subtract it from the total spectrum with the comb incident. The residual pump light from the few-cycle driving pulse is filtered using a combination of 3-mm thick germanium windows and longpass interference filters on germanium and silicon substrates. In contrast to the PPLN configuration, an inline $f_{CEO}$ beat note is not observed due to the unfavorable phase-matching conditions in the OP-GaP crystal for the near-infrared. As a result, conventional $f-2f$ interferometry would be required to stabilize the carrier-envelope-offset frequency of the near-infrared frequency comb. The MIR output would still be offset-free in these conditions. 

Although a single poling period was used in the OP-GaP, driving pulses from waveguides with widths 1280~nm and 1620~nm resulted in output spectra centered at 7.5 and 8.5~\textmu{}m, respectively. The supercontinuum generation in the SiN waveguides with different widths provide a versatile method to optimize the pump-pulse spectrum compared to highly-nonlinear fiber. In the OP-GaP, owing to the lower central frequencies in the long-wave infrared, the central peak of the pump at 1.55 \textmu{}m ($\approx$193.4~THz) participates in the nonlinear frequency conversion. Using the shorter-wavelength spectrum, we performed gas-phase spectroscopy of a commonplace refrigerant, difluoroethane (C$_2$H$_4$F$_2$) from 1000--1400~cm$^{-1}$ (7.1--10~\textmu{}m), a spectral range that is 3--4 times broader than present QCL comb measurements \cite{klocke_single-shot_2018}. The measured absorbance showed quantitative agreement with the NIST WebBook database spectrum (\cite{nist_webbook}, Fig.~\ref{fig:lwir}(b), inset). 

In conclusion, we presented a robust technique to obtain broadband 10 GHz frequency combs in the MIR obtained using Er:fiber technology, electro-optics, and picojoule-scale nonlinear optics with integrated photonics platforms. Such sources should prove valuable for applications such as high-speed spectroscopy and imaging of large biological and chemical compounds. Furthermore, in place of the electro-optic comb generator, recently developed solid-state 10-GHz-rate near-infrared frequency combs \cite{mayer_watt-level_2017} can serve as efficient pump sources for the photonic-chip enabled nonlinear frequency conversion processes that would further simplify the experimental setup and provide robust performance. While state-of-the-art QCLs can provide much higher powers in the mid-infrared, we anticipate that recent advances in nanophotonic lithium niobate \cite{jankowski2020ultrabroadband} and gallium phosphide platforms \cite{wilson2020integrated} combined with careful pulse-shaping of the NIR driver \cite{campo2017shaping, lind2020mid} would further improve the frequency conversion efficiency into the MIR. Such powers can enable applications such as remote sensing using heterodyne interferometry \cite{kostiuk1983remote}.

\noindent\textbf{Acknowledgements} This work was supported by DARPA SCOUT, AFOSR (FA9550-16-1-0016), and NIST. We thank D. Lesko, N. Nader, and P. Acedo for helpful comments.

\noindent\textbf{Disclosure} DRC is a co-founder of Octave Photonics, a company specializing in nonlinear integrated photonics. Certain commercial equipment, instruments, or materials are identified in this paper to foster understanding. Such identification does not imply recommendation or endorsement by the National Institute of Standards and Technology, nor does it imply that the materials or equipment identified are necessarily the best available for the purpose.

\bibliography{sample}

\begin{thebibliography}{10}
\newcommand{\enquote}[1]{``#1''}

\bibitem{cossel_gas-phase_2017}
K.~C. Cossel, E.~M. Waxman, I.~A. Finneran, G.~A. Blake, J.~Ye, and N.~R.
  Newbury, \enquote{Gas-phase broadband spectroscopy using active sources:
  progress, status, and applications [{Invited}],} {\protect\JournalTitle{JOSA
  B}} \textbf{34}, 104--129 (2017).

\bibitem{ycas_high-coherence_2018}
G.~Ycas, F.~R. Giorgetta, E.~Baumann, I.~Coddington, D.~Herman, S.~A. Diddams,
  and N.~R. Newbury, \enquote{High-coherence mid-infrared dual-comb
  spectroscopy spanning 2.6 to 5.2 um,} {\protect\JournalTitle{Nature
  Photonics}} \textbf{12}, 202--208 (2018).

\bibitem{muraviev_massively_2018}
A.~V. Muraviev, V.~O. Smolski, Z.~E. Loparo, and K.~L. Vodopyanov,
  \enquote{Massively parallel sensing of trace molecules and their
  isotopologues with broadband subharmonic mid-infrared frequency combs,}
  {\protect\JournalTitle{Nature Photonics}} \textbf{12}, 209--214 (2018).

\bibitem{Changala2019}
P.~B. Changala, M.~L. Weichman, K.~F. Lee, M.~E. Fermann, and J.~Ye,
  \enquote{Rovibrational quantum state resolution of the c60 fullerene,}
  {\protect\JournalTitle{Science}} \textbf{363}, 49--54 (2019).

\bibitem{fleisher_mid-infrared_2014}
A.~J. Fleisher, B.~J. Bjork, T.~Q. Bui, K.~C. Cossel, M.~Okumura, and J.~Ye,
  \enquote{Mid-{Infrared} {Time}-{Resolved} {Frequency} {Comb} {Spectroscopy}
  of {Transient} {Free} {Radicals},} {\protect\JournalTitle{The Journal of
  Physical Chemistry Letters}} \textbf{5}, 2241--2246 (2014).

\bibitem{klocke_single-shot_2018}
J.~L. Klocke, M.~Mangold, P.~Allmendinger, A.~Hugi, M.~Geiser, P.~Jouy,
  J.~Faist, and T.~Kottke, \enquote{Single-{Shot} {Sub}-microsecond
  {Mid}-infrared {Spectroscopy} on {Protein} {Reactions} with {Quantum}
  {Cascade} {Laser} {Frequency} {Combs},} {\protect\JournalTitle{Analytical
  Chemistry}}  (2018).

\bibitem{Kowligy2019}
A.~S. Kowligy, H.~Timmers, A.~J. Lind, U.~Elu, F.~C. Cruz, P.~G. Schunemann,
  J.~Biegert, and S.~A. Diddams, \enquote{Infrared electric field sampled
  frequency comb spectroscopy,} {\protect\JournalTitle{Science Advances}}
  \textbf{5} (2019).

\bibitem{Weichmann2019}
M.~L. {Weichman}, P.~B. {Changala}, J.~{Ye}, Z.~{Chen}, M.~{Yan}, and
  N.~{Picqu{\'e}}, \enquote{{Broadband molecular spectroscopy with optical
  frequency combs},} {\protect\JournalTitle{Journal of Molecular Spectroscopy}}
  \textbf{355}, 66--78 (2019).

\bibitem{Vasilyev:19}
S.~Vasilyev, I.~Moskalev, V.~Smolski, J.~Peppers, M.~Mirov, V.~Fedorov,
  D.~Martyshkin, S.~Mirov, and V.~Gapontsev, \enquote{Octave-spanning cr:zns
  femtosecond laser with intrinsic nonlinear interferometry,}
  {\protect\JournalTitle{Optica}} \textbf{6}, 126--127 (2019).

\bibitem{Duval:15}
S.~Duval, M.~Bernier, V.~Fortin, J.~Genest, M.~Pich\'{e}, and R.~Vall\'{e}e,
  \enquote{Femtosecond fiber lasers reach the mid-infrared,}
  {\protect\JournalTitle{Optica}} \textbf{2}, 623--626 (2015).

\bibitem{faist_quantum_2016}
J.~Faist, G.~Villares, G.~Scalari, M.~Rosch, C.~Bonzon, A.~Hugi, and M.~Beck,
  \enquote{Quantum {Cascade} {Laser} {Frequency} {Combs},}
  {\protect\JournalTitle{Nanophotonics}} \textbf{5}, 272--291 (2016).

\bibitem{Antipov2016}
S.~Antipov, D.~D. Hudson, A.~Fuerbach, and S.~D. Jackson, \enquote{High-power
  mid-infrared femtosecond fiber laser in the water vapor transmission window,}
  {\protect\JournalTitle{Optica}} \textbf{3}, 1373--1376 (2016).

\bibitem{schliesser_mid-infrared_2012}
A.~Schliesser, N.~Picqué, and T.~W. Hänsch, \enquote{Mid-infrared frequency
  combs,} {\protect\JournalTitle{Nature Photonics}} \textbf{6}, 440--449
  (2012).

\bibitem{erny_mid-infrared_2007}
C.~Erny, K.~Moutzouris, J.~Biegert, D.~Kühlke, F.~Adler, A.~Leitenstorfer, and
  U.~Keller, \enquote{Mid-infrared difference-frequency generation of
  ultrashort pulses tunable between 3.2 and 4.8 um from a compact fiber
  source,} {\protect\JournalTitle{Optics Letters}} \textbf{32}, 1138--1140
  (2007).

\bibitem{Gambetta:08}
A.~Gambetta, R.~Ramponi, and M.~Marangoni, \enquote{Mid-infrared optical combs
  from a compact amplified er-doped fiber oscillator,}
  {\protect\JournalTitle{Opt. Lett.}} \textbf{33}, 2671--2673 (2008).

\bibitem{sell_field-resolved_2008}
A.~Sell, R.~Scheu, A.~Leitenstorfer, and R.~Huber, \enquote{Field-resolved
  detection of phase-locked infrared transients from a compact {Er}:fiber
  system tunable between 55 and 107 {THz},} {\protect\JournalTitle{Applied
  Physics Letters}} \textbf{93}, 251107 (2008).

\bibitem{cruz_mid-infrared_2015}
F.~C. Cruz, D.~L. Maser, T.~Johnson, G.~Ycas, A.~Klose, F.~R. Giorgetta,
  I.~Coddington, and S.~A. Diddams, \enquote{Mid-infrared optical frequency
  combs based on difference frequency generation for molecular spectroscopy,}
  {\protect\JournalTitle{Optics Express}} \textbf{23}, 26814--26824 (2015).

\bibitem{Adler2009}
F.~Adler, K.~C. Cossel, M.~J. Thorpe, I.~Hartl, M.~E. Fermann, and J.~Ye,
  \enquote{Phase-stabilized, 1.5 w frequency comb at 2.8--4.8 $\mu$m,}
  {\protect\JournalTitle{Opt. Lett.}} \textbf{34}, 1330--1332 (2009).

\bibitem{Leindecker:11}
N.~Leindecker, A.~Marandi, R.~L. Byer, and K.~L. Vodopyanov, \enquote{Broadband
  degenerate opo for mid-infrared frequency comb generation,}
  {\protect\JournalTitle{Opt. Express}} \textbf{19}, 6296--6302 (2011).

\bibitem{Marandi:16}
A.~Marandi, K.~A. Ingold, M.~Jankowski, and R.~L. Byer, \enquote{Cascaded
  half-harmonic generation of femtosecond frequency combs in the mid-infrared,}
  {\protect\JournalTitle{Optica}} \textbf{3}, 324--327 (2016).

\bibitem{vainio_mid-infrared_2016}
M.~Vainio and L.~Halonen, \enquote{Mid-infrared optical parametric oscillators
  and frequency combs for molecular spectroscopy,}
  {\protect\JournalTitle{Physical Chemistry Chemical Physics}} \textbf{18},
  4266--4294 (2016).

\bibitem{maidment_long-wave_2018}
L.~Maidment, O.~Kara, P.~G. Schunemann, J.~Piper, K.~McEwan, and D.~T. Reid,
  \enquote{Long-wave infrared generation from femtosecond and picosecond
  optical parametric oscillators based on orientation-patterned gallium
  phosphide,} {\protect\JournalTitle{Applied Physics B}} \textbf{124}, 143
  (2018).

\bibitem{mirmicrocombs16}
M.~Yu, Y.~Okawachi, A.~G. Griffith, M.~Lipson, and A.~L. Gaeta,
  \enquote{Mode-locked mid-infrared frequency combs in a silicon
  microresonator,} {\protect\JournalTitle{Optica}} \textbf{3}, 854--860 (2016).

\bibitem{wang_mid-infrared_2013}
C.~Y. Wang, T.~Herr, P.~Del’Haye, A.~Schliesser, J.~Hofer, R.~Holzwarth,
  T.~W. Hänsch, N.~Picqué, and T.~J. Kippenberg, \enquote{Mid-infrared
  optical frequency combs at 2.5 um based on crystalline microresonators,}
  {\protect\JournalTitle{Nature Communications}} \textbf{4}, ncomms2335 (2013).

\bibitem{petersen_mid-infrared_2014}
C.~R. Petersen, U.~Møller, I.~Kubat, B.~Zhou, S.~Dupont, J.~Ramsay, T.~Benson,
  S.~Sujecki, N.~Abdel-Moneim, Z.~Tang, D.~Furniss, A.~Seddon, and O.~Bang,
  \enquote{Mid-infrared supercontinuum covering the 1.4–13.3 um molecular
  fingerprint region using ultra-high na chalcogenide step-index fibre,}
  {\protect\JournalTitle{Nature Photonics}} \textbf{8}, 830--834 (2014).

\bibitem{kuyken_-chip_2011}
B.~Kuyken, S.~Clemmen, S.~K. Selvaraja, W.~Bogaerts, D.~V. Thourhout,
  P.~Emplit, S.~Massar, G.~Roelkens, and R.~Baets, \enquote{On-chip parametric
  amplification with 26.5 {dB} gain at telecommunication wavelengths using
  {CMOS}-compatible hydrogenated amorphous silicon waveguides,}
  {\protect\JournalTitle{Optics Letters}} \textbf{36}, 552--554 (2011).

\bibitem{Hudson:17}
D.~D. Hudson, S.~Antipov, L.~Li, I.~Alamgir, T.~Hu, M.~E. Amraoui,
  Y.~Messaddeq, M.~Rochette, S.~D. Jackson, and A.~Fuerbach, \enquote{Toward
  all-fiber supercontinuum spanning the mid-infrared,}
  {\protect\JournalTitle{Optica}} \textbf{4}, 1163--1166 (2017).

\bibitem{Nader:19}
N.~Nader, A.~Kowligy, J.~Chiles, E.~J. Stanton, H.~Timmers, A.~J. Lind, F.~C.
  Cruz, D.~M.~B. Lesko, K.~A. Briggman, S.~W. Nam, S.~A. Diddams, and R.~P.
  Mirin, \enquote{Infrared frequency comb generation and spectroscopy with
  suspended silicon nanophotonic waveguides,} {\protect\JournalTitle{Optica}}
  \textbf{6}, 1269--1276 (2019).

\bibitem{huber2000generation}
R.~Huber, A.~Brodschelm, F.~Tauser, and A.~Leitenstorfer, \enquote{Generation
  and field-resolved detection of femtosecond electromagnetic pulses tunable up
  to 41 thz,} {\protect\JournalTitle{Applied Physics Letters}} \textbf{76},
  3191--3193 (2000).

\bibitem{keilmann_time-domain_2004}
F.~Keilmann, C.~Gohle, and R.~Holzwarth, \enquote{Time-domain mid-infrared
  frequency-comb spectrometer,} {\protect\JournalTitle{Optics Letters}}
  \textbf{29}, 1542--1544 (2004).

\bibitem{pupeza_high-power_2015}
I.~Pupeza, D.~Sánchez, J.~Zhang, N.~Lilienfein, M.~Seidel, N.~Karpowicz,
  T.~Paasch-Colberg, I.~Znakovskaya, M.~Pescher, W.~Schweinberger, V.~Pervak,
  E.~Fill, O.~Pronin, Z.~Wei, F.~Krausz, A.~Apolonski, and J.~Biegert,
  \enquote{High-power sub-two-cycle mid-infrared pulses at 100 {MHz} repetition
  rate,} {\protect\JournalTitle{Nature Photonics}} \textbf{9}, 721--724 (2015).

\bibitem{butler2019watt}
T.~Butler, D.~Gerz, C.~Hofer, J.~Xu, C.~Gaida, T.~Heuermann, M.~Gebhardt,
  L.~Vamos, W.~Schweinberger, J.~Gessner \emph{et~al.}, \enquote{Watt-scale
  50-mhz source of single-cycle waveform-stable pulses in the molecular
  fingerprint region,} {\protect\JournalTitle{Optics letters}} \textbf{44},
  1730--1733 (2019).

\bibitem{timmers_molecular_2018}
H.~Timmers, A.~Kowligy, A.~Lind, F.~C. Cruz, N.~Nader, M.~Silfies, G.~Ycas,
  T.~K. Allison, P.~G. Schunemann, S.~B. Papp, and S.~A. Diddams,
  \enquote{Molecular fingerprinting with bright, broadband infrared frequency
  combs,} {\protect\JournalTitle{Optica}} \textbf{5}, 727--732 (2018).

\bibitem{seidel_multi-watt_2018}
M.~Seidel, X.~Xiao, S.~A. Hussain, G.~Arisholm, A.~Hartung, K.~T. Zawilski,
  P.~G. Schunemann, F.~Habel, M.~Trubetskov, V.~Pervak, O.~Pronin, and
  F.~Krausz, \enquote{Multi-watt, multi-octave, mid-infrared femtosecond
  source,} {\protect\JournalTitle{Science Advances}} \textbf{4}, eaaq1526
  (2018).

\bibitem{tu2016stain}
H.~Tu, Y.~Liu, D.~Turchinovich, M.~Marjanovic, J.~K. Lyngs{\o},
  J.~L{\ae}gsgaard, E.~J. Chaney, Y.~Zhao, S.~You, W.~L. Wilson \emph{et~al.},
  \enquote{Stain-free histopathology by programmable supercontinuum pulses,}
  {\protect\JournalTitle{Nature photonics}} \textbf{10}, 534 (2016).

\bibitem{hermes2018mid}
M.~Hermes, R.~B. Morrish, L.~Huot, L.~Meng, S.~Junaid, J.~Tomko, G.~Lloyd,
  W.~Masselink, P.~Tidemand-Lichtenberg, C.~Pedersen \emph{et~al.},
  \enquote{Mid-ir hyperspectral imaging for label-free histopathology and
  cytology,} {\protect\JournalTitle{Journal of Optics}} \textbf{20}, 023002
  (2018).

\bibitem{seddon2018prospective}
A.~B. Seddon, B.~Napier, I.~Lindsay, S.~Lamrini, P.~M. Moselund, N.~Stone,
  O.~Bang, and M.~Farries, \enquote{Prospective on using fibre mid-infrared
  supercontinuum laser sources for in vivo spectral discrimination of disease,}
  {\protect\JournalTitle{Analyst}} \textbf{143}, 5874--5887 (2018).

\bibitem{muller_quantum_2018}
E.~A. Muller, B.~Pollard, H.~A. Bechtel, R.~Adato, D.~Etezadi, H.~Altug, and
  M.~B. Raschke, \enquote{Nanoimaging and control of molecular vibrations
  through electromagnetically induced scattering reaching the strong coupling
  regime,} {\protect\JournalTitle{ACS Photonics}} \textbf{5}, 3594--3600
  (2018).

\bibitem{carlson_ultrafast_2017}
D.~R. {Carlson}, D.~D. {Hickstein}, W.~{Zhang}, A.~J. {Metcalf}, F.~{Quinlan},
  S.~A. {Diddams}, and S.~B. {Papp}, \enquote{{Ultrafast electro-optic light
  with subcycle control},} {\protect\JournalTitle{Science}} \textbf{361},
  1358--1363 (2018).

\bibitem{mayer_offset-free_2016}
A.~Mayer, C.~Phillips, C.~Langrock, A.~Klenner, A.~Johnson, K.~Luke,
  Y.~Okawachi, M.~Lipson, A.~Gaeta, M.~Fejer, and U.~Keller,
  \enquote{Offset-{Free} {Gigahertz} {Midinfrared} {Frequency} {Comb} {Based}
  on {Optical} {Parametric} {Amplification} in a {Periodically} {Poled}
  {Lithium} {Niobate} {Waveguide},} {\protect\JournalTitle{Physical Review
  Applied}} \textbf{6}, 054009 (2016).

\bibitem{Rockmore:19}
R.~Rockmore, A.~Laurain, J.~V. Moloney, and R.~J. Jones, \enquote{Offset-free
  mid-infrared frequency comb based on a mode-locked semiconductor laser,}
  {\protect\JournalTitle{Opt. Lett.}} \textbf{44}, 1797--1800 (2019).

\bibitem{yu2018silicon}
M.~Yu, Y.~Okawachi, A.~G. Griffith, N.~Picqu{\'e}, M.~Lipson, and A.~L. Gaeta,
  \enquote{Silicon-chip-based mid-infrared dual-comb spectroscopy,}
  {\protect\JournalTitle{Nature Communications}} \textbf{9} (2018).

\bibitem{scalari2019chip}
G.~Scalari, J.~Faist, and N.~Picqu{\'e}, \enquote{On-chip mid-infrared and thz
  frequency combs for spectroscopy,} {\protect\JournalTitle{Applied Physics
  Letters}} \textbf{114} (2019).

\bibitem{Yan2017}
M.~{Yan}, P.-L. {Luo}, K.~{Iwakuni}, G.~{Millot}, T.~W. {H{\"a}nsch}, and
  N.~{Picqu{\'e}}, \enquote{{Mid-infrared dual-comb spectroscopy with
  electro-optic modulators},} {\protect\JournalTitle{Light: Science \&
  Applications}} \textbf{6}, e17076 (2017).

\bibitem{Jerez2018}
B.~Jerez, P.~MartÃ­n-Mateos, F.~Walla, C.~de~Dios, and P.~Acedo,
  \enquote{Flexible electro-optic, single-crystal difference frequency
  generation architecture for ultrafast mid-infrared dual-comb spectroscopy,}
  {\protect\JournalTitle{ACS Photonics}} \textbf{5}, 2348--2353 (2018).

\bibitem{coddington_dual-comb_2016}
I.~Coddington, N.~Newbury, and W.~Swann, \enquote{Dual-comb spectroscopy,}
  {\protect\JournalTitle{Optica}} \textbf{3}, 414--426 (2016).

\bibitem{carlson2019generating}
D.~R. Carlson, P.~Hutchison, D.~D. Hickstein, and S.~B. Papp,
  \enquote{Generating few-cycle pulses with integrated nonlinear photonics,}
  {\protect\JournalTitle{Optics Express}} \textbf{27}, 37374--37382 (2019).

\bibitem{asobe_high-power_2008}
M.~Asobe, O.~Tadanaga, T.~Yanagawa, T.~Umeki, Y.~Nishida, and H.~Suzuki,
  \enquote{High-power mid-infrared wavelength generation using difference
  frequency generation in damage-resistant {Zn}:{LiNbO}3 waveguide,}
  {\protect\JournalTitle{Electronics Letters}} \textbf{44}, 288--290 (2008).

\bibitem{iwakuni_generation_2016}
K.~Iwakuni, S.~Okubo, O.~Tadanaga, H.~Inaba, A.~Onae, F.-L. Hong, and
  H.~Sasada, \enquote{Generation of a frequency comb spanning more than 3.6
  octaves from ultraviolet to mid infrared,} {\protect\JournalTitle{Optics
  Letters}} \textbf{41}, 3980--3983 (2016).

\bibitem{nist_webbook}
P.~J. Linstrom and W.~G. Mallard, \enquote{Nist chemistry webbook, nist
  standard reference database number 69,} {\protect\JournalTitle{National
  Institute of Standards and Technology, Gaithersburg, MD 20899}}  (2020).

\bibitem{mayer_watt-level_2017}
A.~S. Mayer, C.~R. Phillips, and U.~Keller, \enquote{Watt-level 10-gigahertz
  solid-state laser enabled by self-defocusing nonlinearities in an
  aperiodically poled crystal,} {\protect\JournalTitle{Nature Communications}}
  \textbf{8}, 1673 (2017).

\bibitem{jankowski2020ultrabroadband}
M.~Jankowski, C.~Langrock, B.~Desiatov, A.~Marandi, C.~Wang, M.~Zhang, C.~R.
  Phillips, M.~Lon{\v{c}}ar, and M.~Fejer, \enquote{Ultrabroadband nonlinear
  optics in nanophotonic periodically poled lithium niobate waveguides,}
  {\protect\JournalTitle{Optica}} \textbf{7}, 40--46 (2020).

\bibitem{wilson2020integrated}
D.~J. Wilson, K.~Schneider, S.~H{\"o}nl, M.~Anderson, Y.~Baumgartner,
  L.~Czornomaz, T.~J. Kippenberg, and P.~Seidler, \enquote{Integrated gallium
  phosphide nonlinear photonics,} {\protect\JournalTitle{Nature Photonics}}
  \textbf{14}, 57--62 (2020).

\bibitem{campo2017shaping}
G.~Campo, A.~Leshem, F.~Cappelli, I.~Galli, P.~C. Pastor, A.~Arie,
  P.~De~Natale, and D.~Mazzotti, \enquote{Shaping the spectrum of a
  down-converted mid-infrared frequency comb,} {\protect\JournalTitle{JOSA B}}
  \textbf{34}, 2287--2294 (2017).

\bibitem{lind2020mid}
A.~J. Lind, A.~Kowligy, H.~Timmers, F.~C. Cruz, N.~Nader, M.~C. Silfies, T.~K.
  Allison, and S.~A. Diddams, \enquote{Mid-infrared frequency comb generation
  and spectroscopy with few-cycle pulses and $\chi$ (2) nonlinear optics,}
  {\protect\JournalTitle{Physical Review Letters}} \textbf{124}, 133904 (2020).

\bibitem{kostiuk1983remote}
T.~Kostiuk and M.~J. Mumma, \enquote{Remote sensing by ir heterodyne
  spectroscopy,} {\protect\JournalTitle{Applied optics}} \textbf{22},
  2644--2654 (1983).

\end{thebibliography}

\end{document}